\documentclass[a4paper]{article}

\usepackage{icrc2013}

\title{Dielectric Coatings for IACT Mirrors}

\shorttitle{Dielectric Coatings for IACT Mirrors}

\authors{
A.~F{\"o}rster$^{1}$, 
T.~Armstrong$^{2}$,
P.~Chadwick$^{2}$,
for the CTA Collaboration, and
M.~Held$^{3}$.
}

\afiliations{
$^{1}$  Max-Planck-Institut f{\"u}r Kernphysik, Heidelberg, Germany  \\
$^{2}$  University of Durham, Durham, United Kingdom  \\
$^{3}$  BTE Bedampfungstechnik GmbH, Elsoff, Germany
}

\email{andreas.foerster@mpi-hd.mpg.de}

\abstract{
Imaging Atmospheric Cherenkov Telescopes for
very-high energy gamma-ray astronomy need mirror with high reflectance
roughly in the wavelength between 300 and 550 nm. The current standard
reflective layer of such mirrors is aluminum. 
Being permanently exposed to
the environment they show a constant degradation over the years.
New and improved dielectric coatings have been developed to enhance
their resistance to environmental impact and to extend their possible
lifetime. In addition, these customized coatings have an increased   
reflectance of over 95\% and are designed to significantly lower      
the night-sky background contribution. The development of such 
coatings for mirrors with
areas up to 2~m$^2$ and low application temperatures to suite the
composite materials used for the new mirror susbtrates
of the Cherenkov Telescope Array (CTA) and the
results of extensive durability tests are presented.
}


\begin{document}
\maketitle

%Begin a section.
\section{Introduction}

Imaging Atmospheric Cherenkov Telescopes (IACTs) for very-high 
energy (VHE) gamma-ray astronomy image the Cherenkov light of particle 
showers in the atmosphere onto a photosensitive detector. 
The wavelength range of interest is roughly between 300 and 550~nm. 
Typically, IACTs have tesselated mirror areas of the order of 1~m$^2$ and 
larger. The current standard  (e.g. in H.E.S.S., VERITAS 
and partially MAGIC) is mirrors with glass surfaces, coated on 
the front surface with aluminium (Al) which is protected by  
a single protective layer (e.g. SiO$_2$, Al$_2$O$_3$). Not being 
protected by a dome, the mirrors are constantly exposed to the 
environment and show a loss of reflectance of a few per cent per year. 
This requires re-coating of all mirrors after a few years of operation. 
For the 
future CTA observatory (see \cite{Hofmann:2010})  
with a total planned mirror area of about
10,000 m$^2$ this would 
mean a significant maintenance effort. Coatings which increase the 
lifetime of the mirrors can therefore play a major role in
keeping the maintenance costs of the observatory low. 
In addition, coatings with higher reflectance in the relevant wavelength 
range compared to the classical Al + SiO$_2$ coatings 
will increase the sensitivity of the instrument while a reduced 
reflectance above roughly 550 nm can help to suppress sensitivity 
to background light
from the night sky.

\section{Coatings under Investigation}

Aluminium coatings with a single SiO$_2$ layer typically show 
a reflectance of 80 to 90\% between 300 and 550~nm.
To improve the reflectance and the durability of the 
mirrors two commercially available coating options are currently
under investigation in this study (further coating designs 
are being investigated at the University of 
T\"ubingen and are presented in~\cite{Bonardi:2013}.).

\begin{figure}[!h]
  \begin{center}
    \includegraphics[width = 1.1\hsize]{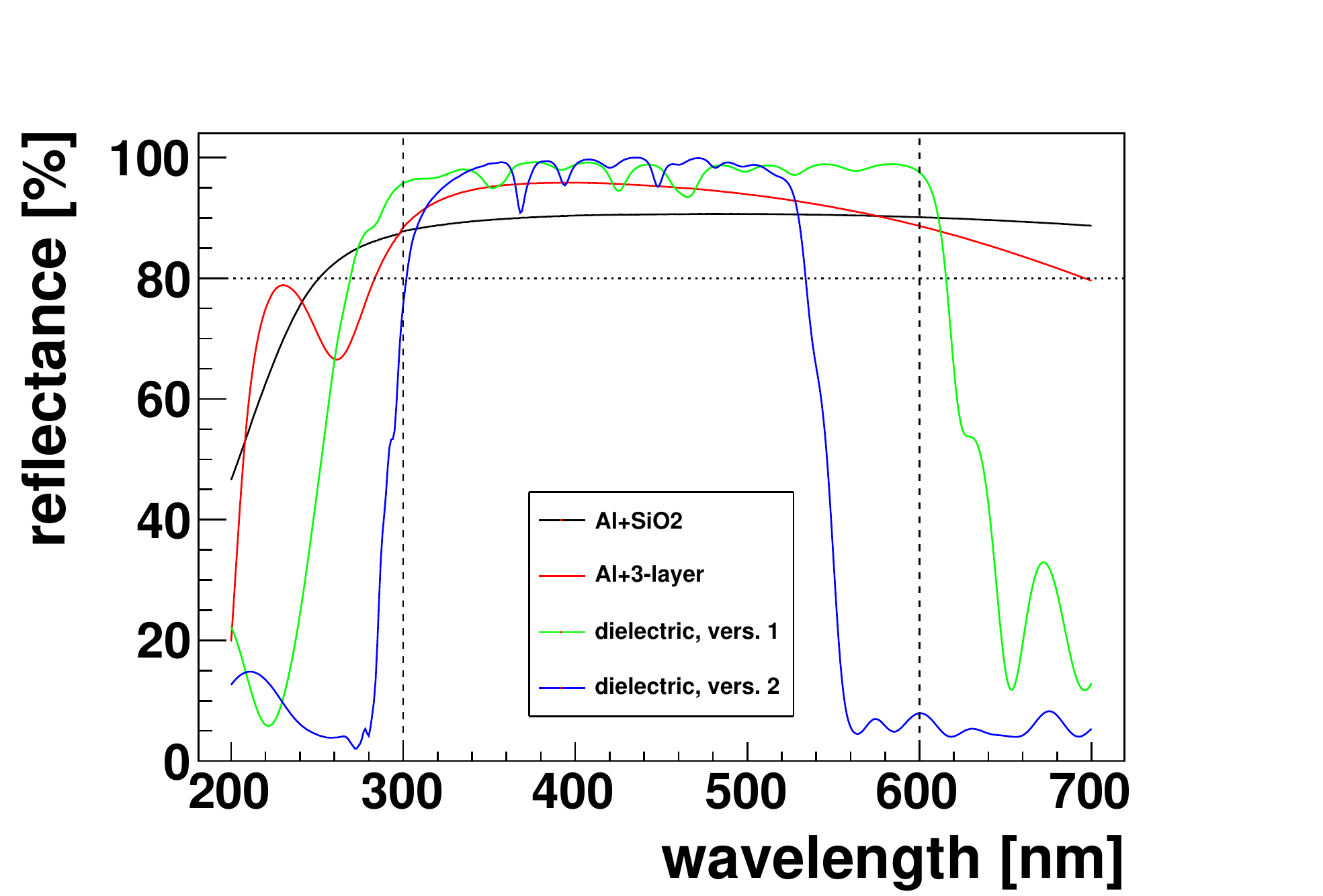} 
    \caption{Comparsion of the spectral reflectance of the two newly
investigated coatings (Al + SiO$_2$ + HfO$_2$ + SiO$_2$ and dielectric) 
compared to Al + SiO$_2$.}
    \label{fig:fig001}
  \end{center}
\end{figure}

(a) A three-layer protective coating (SiO$_2$ + HfO$_2$ + SiO$_2$) on top 
of an Al coating. This already enhances the reflectance by 
about 5\%. Fig.~\ref{fig:fig001} shows the reflectance 
of this coating in comparison 
to the reflectance of an Al + SiO$_2$ coating.

(b) A dielectric coating, consisting of a stack of many 
alternating layers of two materials with different refractive 
indices, without any metallic layer. Properly optimized, this
results into an pure interference mirror that allows 
a box-shaped reflectance curve to be custom-made with $>95\%$ reflectance
in a defined wavelength range and $<30\%$ elsewhere. 
The first attempts were designed such that a range of 300 to 600~nm
was covered. This is called ``version 1'' in the following.
For ``version 2'' the design was adjusted such that
a cut-off around 550~nm allows the reduction of the 
the night-sky background contribution (first emission line around 556~nm). 
The latter might become important in combination with a possible
future replacement of the current photomultiplier tubes in the
photodetectors of the telescopes (which
are not particularly susceptible to night-sky background) 
by silicon  detectors
that have a good quantum efficiency for wavelengths above
550~nm. 
The reflectance curves of these coatings are as well shown
in Fig.~\ref{fig:fig001}.

Most durability testing reported on in the following has been performed 
on version 1 and is currently ongoing for version 2. 
Nevertheless, the materials and the technology used are exactly the
same for both versions, the only difference is the number of layers.

\section{Application to Large Substrates and at Low Temperatures}

\begin{figure}[!h]
  \begin{center}
    \includegraphics[width = 1.1\hsize]{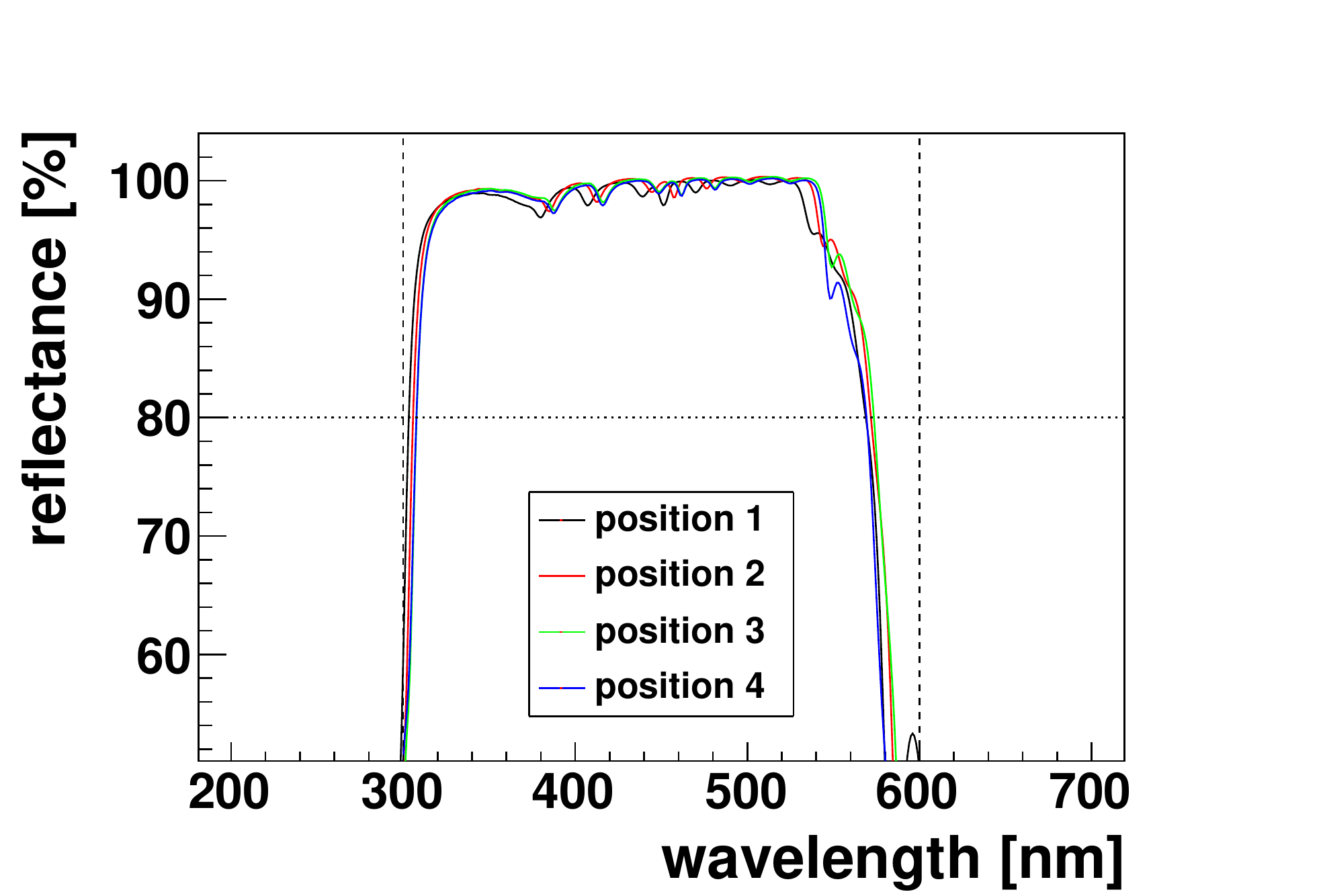} 
    \caption{Reflectance of the dielectric coating measured at 
4 points distributed over the diameter of a 2~m$^2$ hexagonal mirror.}
    \label{fig:fig002}
  \end{center}
\end{figure}

The largest mirrors needed for CTA will be of hexagonal shape with
1.5~m distance from flat to flat, resulting in a mirror area 
of roughly 2~m$^2$. 
While the Al based coatings are in principle 
commercially available for substrates of this size, applying the dielectric
coatings to such large surfaces was more challenging and needed additional 
development effort. Given that these are interference coatings consisting
of many layers of materials with different refractive indices a 
homogeneuos thickness of each layer over the full mirror size is important.
Fig.~\ref{fig:fig002} shows the uniformity achieved in the reflectance
over the diameter of such a big mirror.

The second challenge was created by the mirror substrates themselves. 
Many substrate technologies under development for CTA are sandwich 
structures of different materials that are glued 
together~\cite{Canestrari:2013,Brun:2013, Foerster:2013}. Most 
of these glues cannot be heated to temperatures above 80$^{\circ}$~C
without damage. While Al-based coatings can be applied without 
heating the substrate and with only limited heat impact from the evaporation
sources during the coating process, a special process was
developed for the dielectric coating to keep the substrate temperature
below the required limit. This comes at the expense of longer coating times
and therefore higher costs. In parallel, the option is being investigated
to coat only the front glass sheet and then to construct the sandwich 
with this coated sheet rather than coating the final mirror. This way the
the more costly low-temperature process would not be needed.

\section{Durability Testing}

A series of durability tests have been performed in the laboratory
with small glass samples coated with the different coatings
to evaluate their resistance to environmental impact. \\

\begin{figure}[!h]
  \begin{center}
    \includegraphics[width = 0.8\hsize, angle = +90]{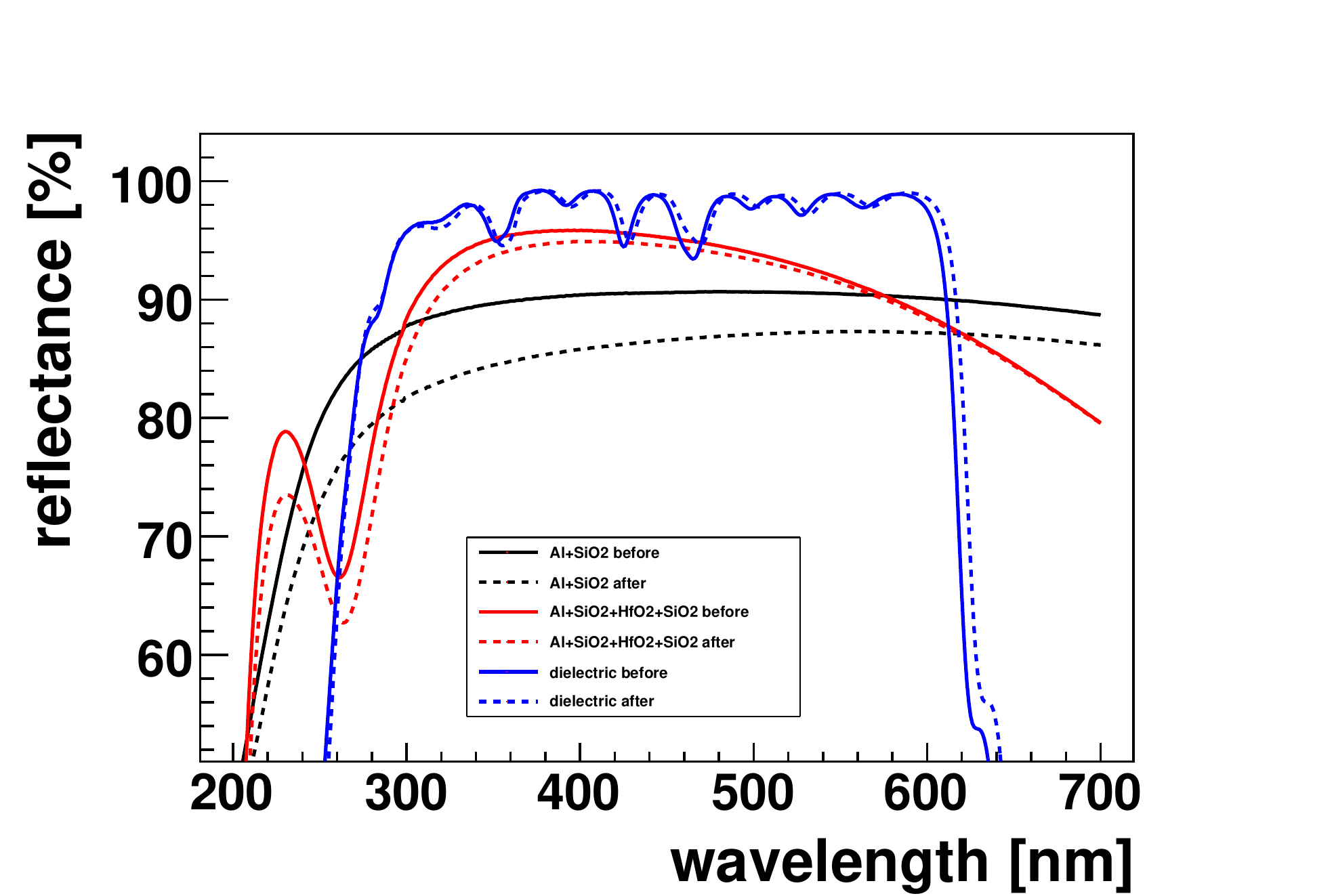} 
    \caption{Comparsion of the spectral reflectance before (full lines) and after (dashed lines) temperature and humidity cycling for all three coatings.}
    \label{fig:fig003}
  \end{center}
\end{figure}

{\bf Temperature and humidity cycling:} Samples of the 
three-layer coating as well as of version 1 of the dielectric coating
have been exposed to overlapping cycles in temperature 
(-10$^{\circ}$~C $<$ T $<$ 60$^{\circ}$~C; 5 h cycle duration) and 
in humidity 
(5\% to 95\%;  8h cycle duration) for a total of approximately 8000~h.
The different cycle duration was chosen to expose the samples
to all possible combinations of temperature and humidity.
The reflectance as a function of wavelength 
of the samples has been measured with a spectrophotometer
(angle of incidence 7$^{\circ}$) before and after
the cycling. The results of these measurements are shown 
in Fig.~\ref{fig:fig003}.

The classical Al + SiO$_2$ coating shows a significant loss of reflectance.
The Al coating with the three layer protective coating exhibts a much 
smaller change in reflectance after the cycling and the dielectric
coating has not changed its reflective properties at all within
the accuracy of the measurement. Samples of version 2 of the 
dielectric coating optimized for the cutoff at 550~nm are currently 
undergoing the same test.
\\

{\bf Adhesion tests:} Coating adhesion was tested by applying a clear tape 
with an adhesion power of 16~N per 25~mm to
the coated surface of the samples and removing it at a rate of $<1$~s per
25~mm (so-called ``snap test''). All tested samples survived this procedure
without any removal of coating, including the new coatings under
investigation and the classical Al+SiO$_2$ coating.
\\

\begin{figure}[!h]
  \begin{center}
    \includegraphics[width = 0.8\hsize]{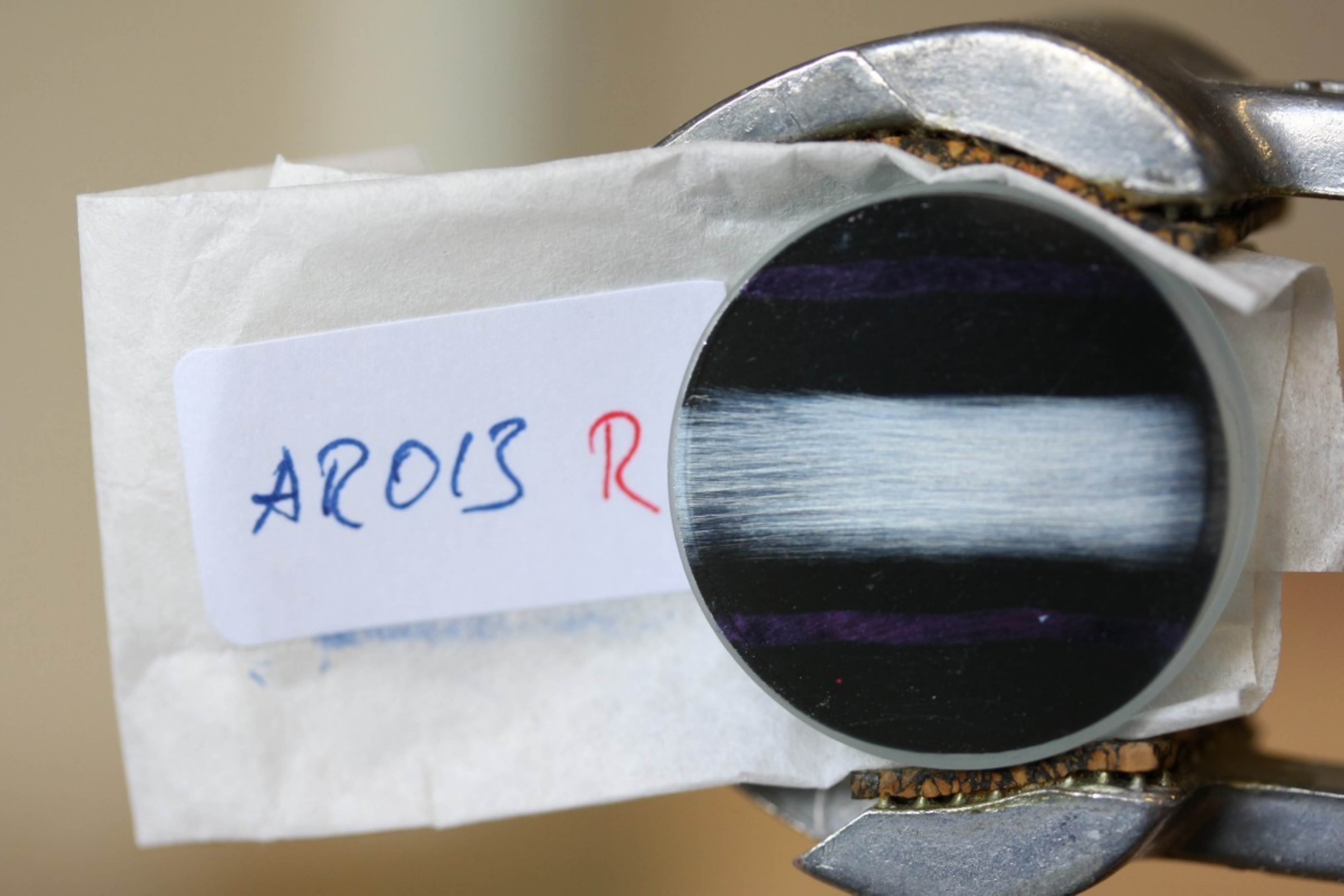} 
    \includegraphics[width = 0.8\hsize]{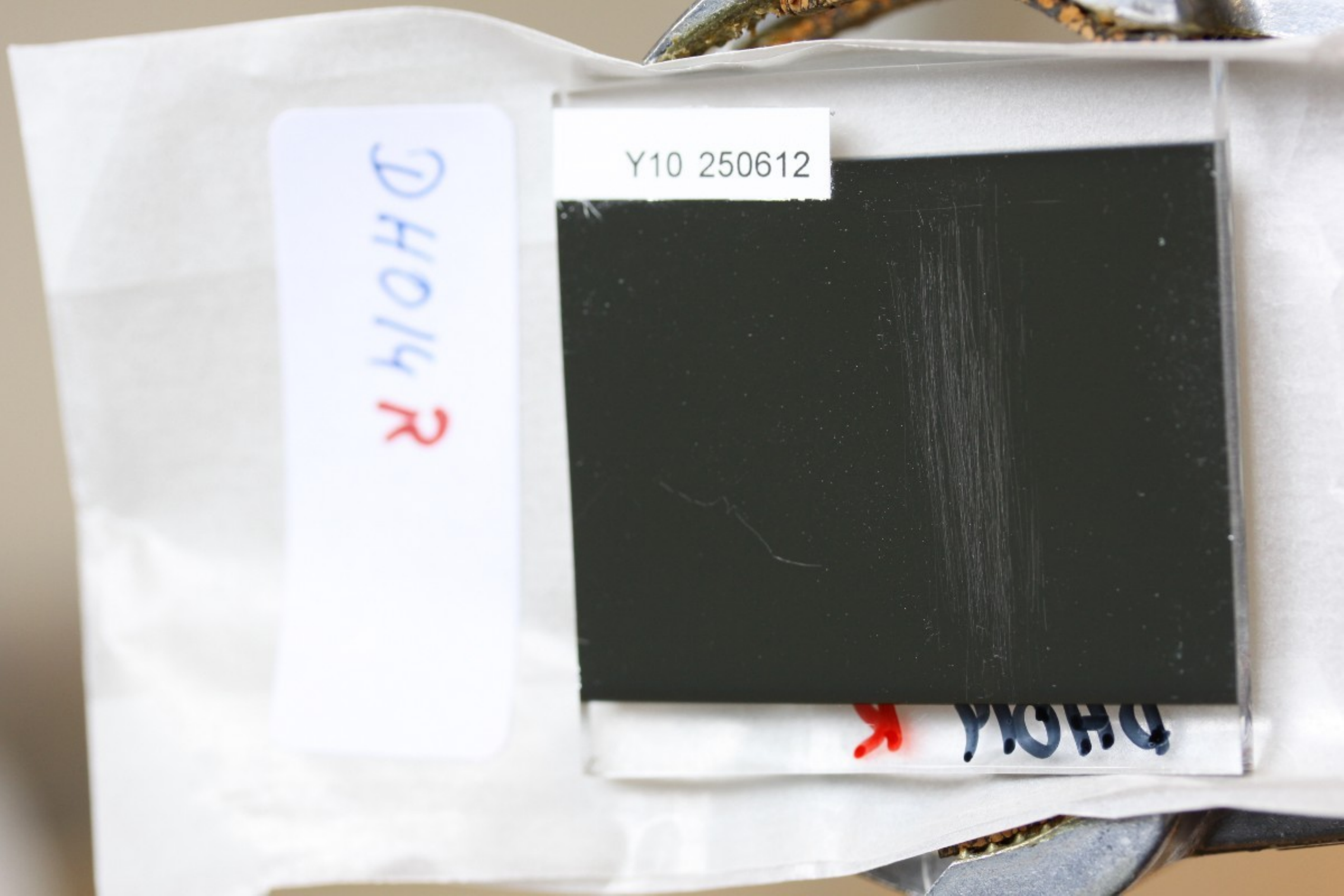} 
    \includegraphics[width = 0.8\hsize]{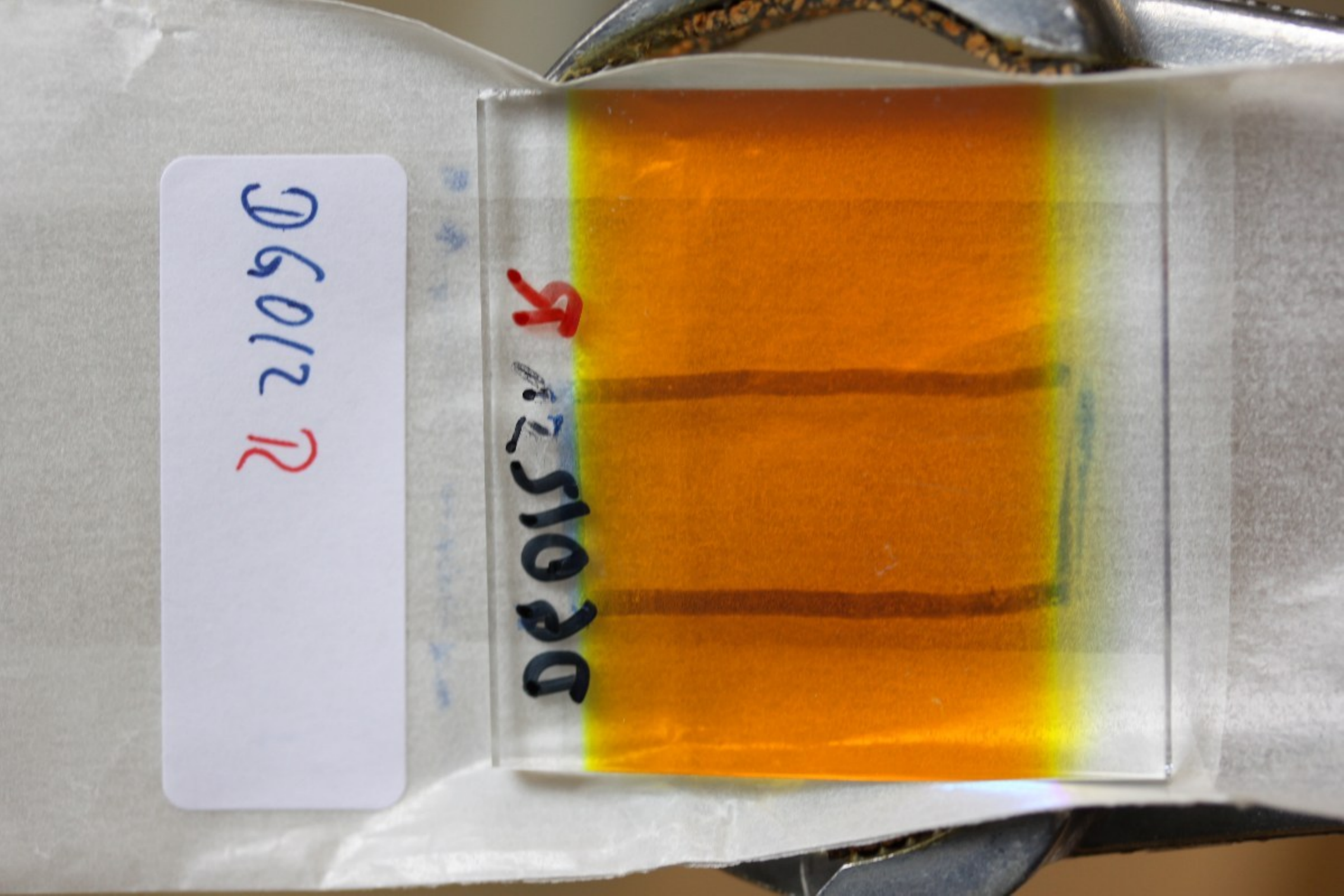} 
    \caption{Coating samples after the severe abrasion test. 
             Upper:Al + SiO$_2$.
             Middle: Al + SiO$_2$ + HfO$_2$ + SiO$_2$. 
             Lower: dielectric coating.}
    \label{fig:fig004}
  \end{center}
\end{figure}

\begin{figure}[!h]
  \begin{center}
    \includegraphics[width = 0.8\hsize]{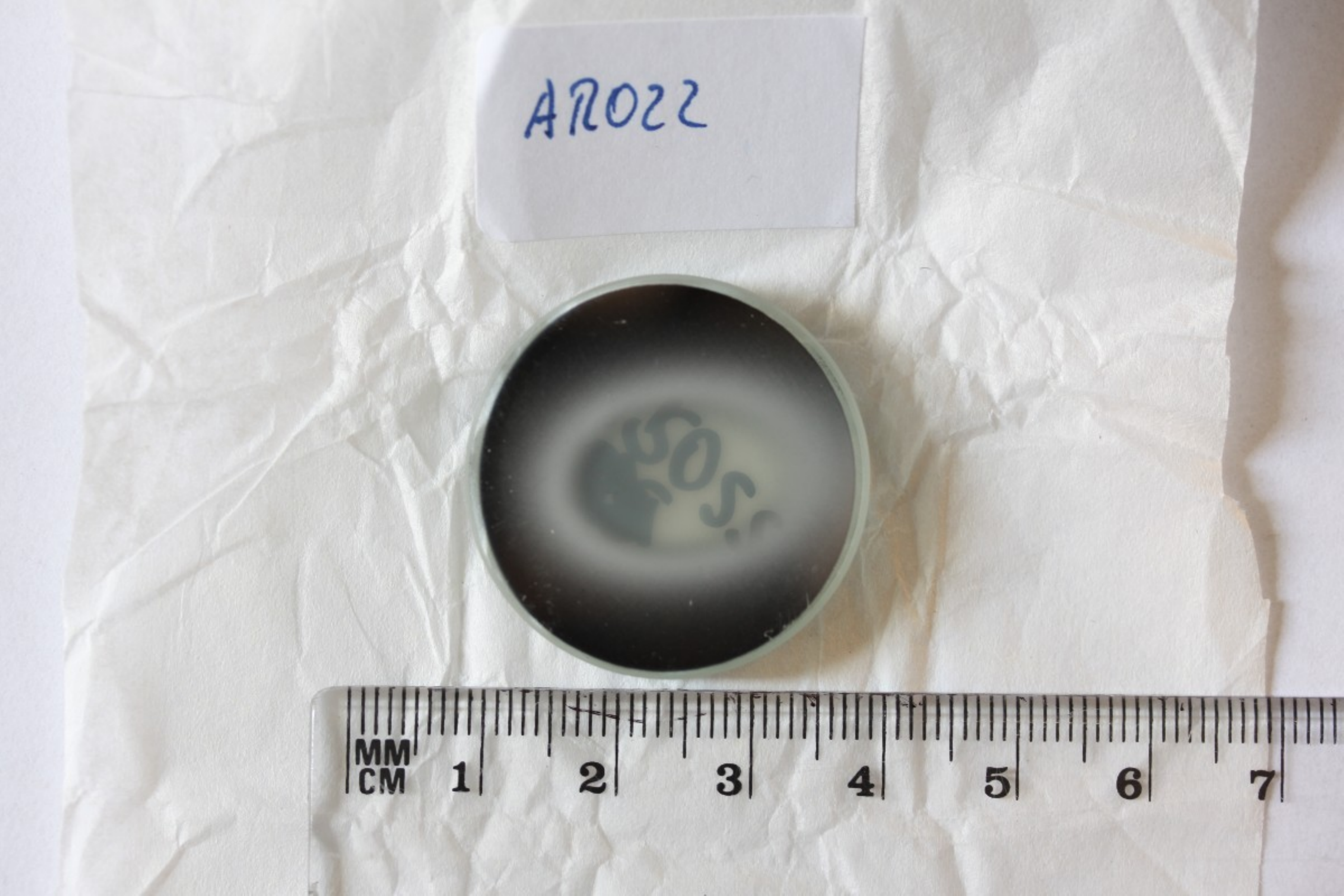} 
    \includegraphics[width = 0.8\hsize]{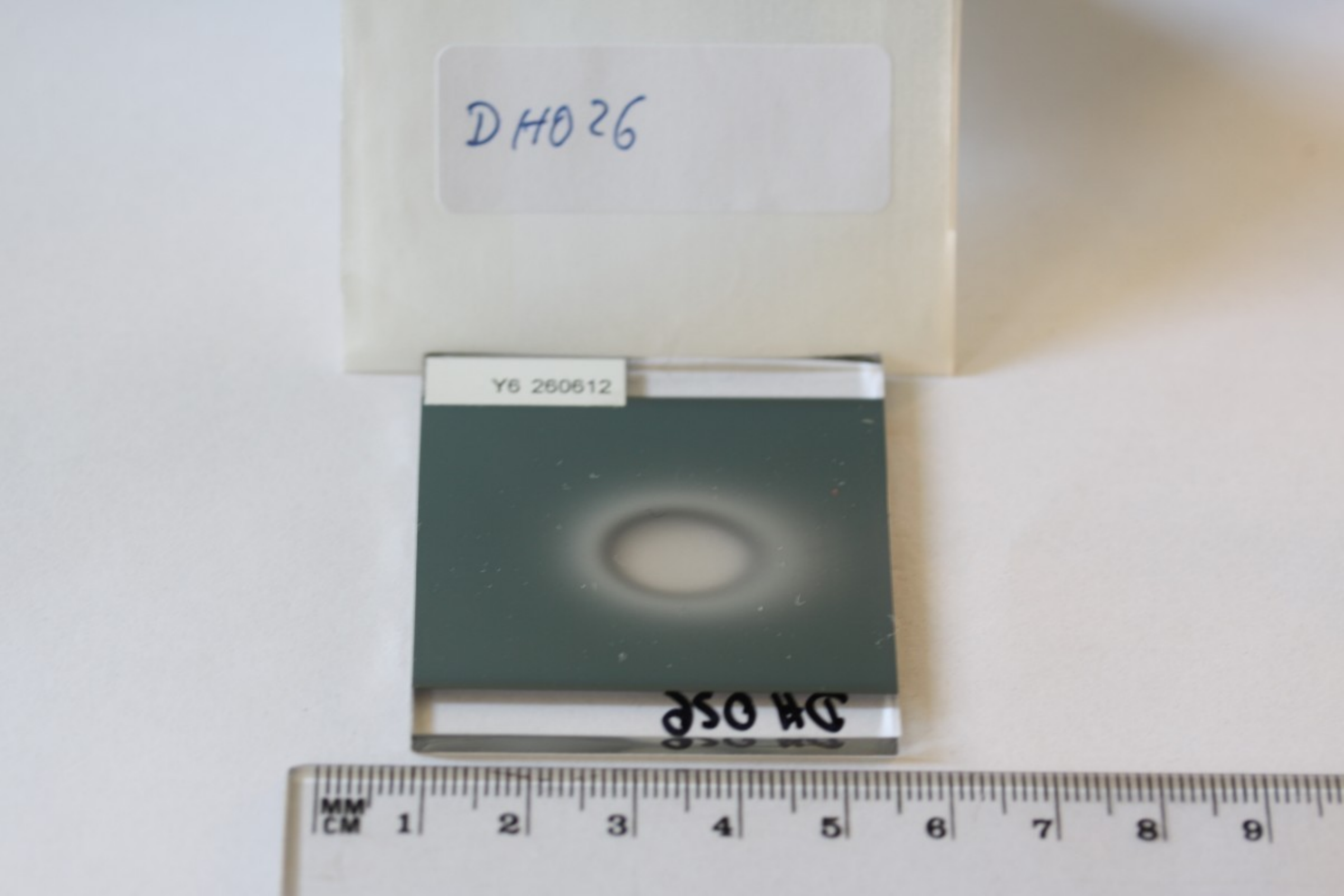} 
    \includegraphics[width = 0.8\hsize]{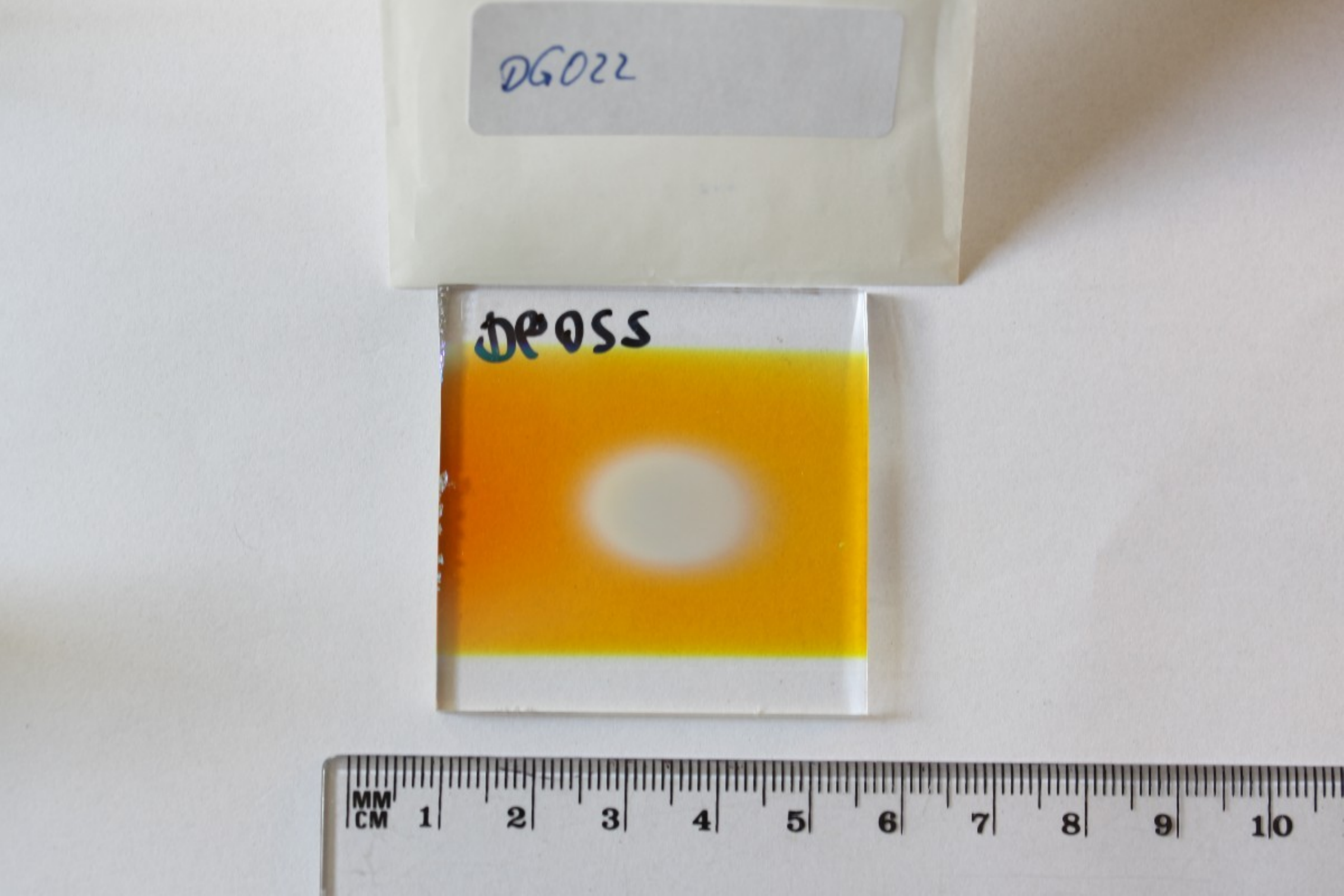} 
    \caption{Coating samples after the sand-blasting test. Upper:Al + SiO$_2$.
             Middle: Al + SiO$_2$ + HfO$_2$ + SiO$_2$. 
             Lower: dielectric coating.}
    \label{fig:fig005}
  \end{center}
\end{figure}

{\bf Abrasion tests:} Three different abrasion tests have been 
performed on samples with all three coatings:

a) A standard cheesecloth test using a force of 
10~N and 50 strokes on the coated surface was performed.
The Al+SiO$_2$ reference samples showed mild to moderate abrasion
under this test (defined as few to many visible scratches 
left behind after the test), the three-layer coating showed 
very mild to mild abrasion (one or two to few scratches) and the
dielectric coating none to very mild abrasion (zero to one or two scratches).
\\

b) In a more severe test an eraser was used to perform 20 strokes with a force
of 10~N. After this test all three coatings showed
signs of abrasion, but again to very different levels: 
the reference samples coated with SiO$_2$ showed serious to 
severe abrasion (removal of some to most of the coating), 
the three-layer coating moderate to severe abrasion (many scratches
to removal of some coating), and the dielectric coating
only showed very mild to mild abrasion (one or two to few scratches).
Figure~\ref{fig:fig004} shows a few samples after this severe abrasion test.
\\

c) Samples with all three coatings were exposed to a sand-blasting
test. The abrading medium used was silicon carbide with a grade of 
220~$\mu$m. The flow rate was approximately 25~g/min and the total
amount of abrading medium used per sample was 125~g. 
The setup was operated using an air pressure of 15~kPa and the air 
was fed in at a rate of 50~l/min. The sample was placed under an angle
of 45$^{\circ}$ under the abrasive jet nozzle. This test results into
an ellipse on the coated surface on which the coating is partially and/or
fully removed.
The sizes of these ellipses are a measure of how easily the coating is
abraded. Figure~\ref{fig:fig005} shows three samples after the 
sand-blasting test;
at the top the Al + SiO$_2$ coating, in the middle the 
Al + SiO$_2$ + HfO$_2$ + SiO$_2$ coating and at the bottom the
dielectric coating.

Three elliptical areas have been measured to quantify the different 
abrasion levels of the three coatings: 
The ``clear area'', meaning the central region in which the coating has been
fully removed, the darker ``partially clear area'' around it and 
the penumbra being the reach of the silicon carbide, 
which shows as a light `halo'. The results are given in 
Tab.~\ref{table:ellipses} and demonstrate that the abrasion
resistance of the dielectric coating is significantly higher than
of the Al-based coatings.
\begin{table}[!h]
\begin{center} 
\begin{tabular}{|c|c|c|c|}
\hline
Coating & Clear Area & Partially Clear & Penumbra \\
\hline
Al+SiO$_2$ & $103 \pm 7$ & $144 \pm 1$ & $420 \pm 40$ \\
3-layer & $98.9 \pm 0.05$ & $136\pm 4$ & $400 \pm 20$ \\
dielectric & $ 0 $ & $114 \pm 3$ & $320 \pm 10$ \\
\hline
\end{tabular}
\caption{Average areas of abrasion ellipses in mm$^2$ resulting from 
the sand blasting test on the three different types of coatings. Given are
mean values and standard deviations over all tested samples of each 
coating type.}
\label{table:ellipses}
\end{center}
\end{table}
\\

{\bf Artificial Bird Faeces:} Samples of all coatings have been treated with
pancreatin, a pancreas enzyme that is regularly used to simulate the effects
of bird feaces on lacquers and other material. A 1:2 mixture of pancreatin
and de-ionized water was applied to the coated surfaces of the samples
and they were ``baked'' for 4 weeks at 40$^{\circ}$~C in a climate 
chamber to simulate the effect of the bird faeces staying on the mirror
surface for some time in a hot and dry environment as is typical
for locations of Cherenkov telescopes. No influence on either
of the the three coatings was observed after cleaning the samples and 
repeated reflectance measurements.
\\

\section{Conclusions and Outlook}

In the laboratory tests described above, the three layer protective
coating on top of an aluminium coating performs slightly better than 
the standard Al + SiO$_2$ coating. The dielectric coating shows 
a significantly better performance still. 
Nevertheless, the predictive power of 
these laboratory tests for the real outdoor performance is not 
clearly established and additional outdoor experience is needed. 
Over the last 2 years all of the approximately
1520 mirrors of the 4 original telescopes of the H.E.S.S. experiment 
in Namibia have been exchanged and
refurbished. 380 mirrors have the standard Al + SiO$_2$ coating,
rougly 1040 the Al + three layer coating, and about 100 mirrors the dielectric
coating. This way data on long-term outdoor exposure will become available. 

One problem of the three layer 
protective coating directly applied on top of the Al layer has been noted
for these mirrors: it was observed
that in the first months after coating the interference minimum 
around 280 nm (see Figs.~\ref{fig:fig001} and~\ref{fig:fig003}) was getting
deeper, slightly affecting the reflectance at 300~nm as well. To solve this
problem, an additional protection layer inbetween the Al and the 
three-layer coating is now being applied to prevent this effect. Detailed
tests of this improved coating are ongoing.

A significant problem of the dielectric coating has been detected in a study
that intended to compare different substrate technologies in terms
of the probability to form condensation on the reflective surface 
\cite{Chadwick:2013}. It was observed that the same substrates are more likely
to mist over if coated with the dielectric coating rather than with
an Al based coating. Laboratory tests 
have associated this effect with a much higher emissivity of the 
dielectric coatings in the infrared (8-14~$\rm{\mu}$m). 
Investigations are ongoing
to solve this problem by applying an additional layer below the 
dielectric coating that does not reflect in the regime of the
night-sky background but has a high reflectance in the mid- infrared.

To conclude, the Al + SiO$_2$ + HfO$_2$ + SiO$_2$ as well as the 
dielectric coating are now readily available alternatives 
to the standard Al + SiO$_2$ coating. 
The three layer coating provides an about 5\%
better reflectance and a slightly better performance in durability tests
in the laboratory at no significant extra cost. The dielectric coating
can now be applied up to substrate areas of 2~m$^2$ and at substrate 
temperatures $<80^{\circ}$~C during the coating process. 
This covers the largest mirrors
forseen for CTA and is suitable for the application on glued
sandwich substrates. It provides a significantly better 
reflectance in the desired wavelength range,
a significant suppression of the night-sky background and 
a significantly better performances in the durability tests, but it 
needs further improvement concerning the high emissivity in the infrared
that leads to a higher probability of forming condensation on the mirrors.

\section*{Acknowledgements}

We gratefully acknowledge support from the agencies and
organisations listed in this page: 
http://www.cta-observatory.org/?q=node/22.


\begin{thebibliography}{}

\bibitem{Hofmann:2010} W.~Hofmann et al.,
 %"Design Concepts for the Cherenkov Telescope Array"
 arXiv:1008.3702 (2010)

\bibitem{Bonardi:2013} A.~Bonardi et al.,
   these proceedings (2013) contribution 0207

\bibitem{Canestrari:2013} R.~Canestrari et al.,
  Optical Engimeering 52 (2013) 051204

\bibitem{Brun:2013} P.~Brun et al.,
  NIM A 714 (2013) 58

\bibitem{Foerster:2013} A.~F{\"o}rster et al., 
   these proceedings (2013) contribution 0747

\bibitem{Chadwick:2013} P.~Chadwick et al.,
   these proceedings (2013) contribution 0847

%\bibitem{Pareschi:2008} Pareschi, G. et al., 
%  %"Glass mirrors by cold slumping to cover 100~m$^2$ of the MAGIC~II Cherenkov telescope
%  %reflecting surface", 
%  SPIE Proc., 7018, p. 70180W (2008)

%\bibitem{Vernani:2008} Vernani, D. et al., 
%  %"Development of cold-slumping glass mirrors for imaging Cherenkov telescopes ",
%  SPIE Proc., 7018, p. 70180V (2008)  


\end{thebibliography}
\end{document}